\documentclass{sig-alternate}

\usepackage{graphics,epsfig,color,url,algorithm,algorithmic,comment,multirow,cite,mdwlist,amsmath,wrapfig,subfigure,graphicx,version}
\date{}

\includeversion{techreport}
\excludeversion{ndss}

\makeatletter
\let\@copyrightspace\relax
\makeatother

\title{X-Vine: Secure and Pseudonymous Routing Using Social Networks}

\numberofauthors{3} 
\author{
\alignauthor Prateek Mittal\\
       \affaddr{Dept. of ECE}\\
       \affaddr{University of Illinois}\\
       \email{mittal2@illinois.edu}
\alignauthor Matthew Caesar\\
       \affaddr{Dept. of CS}\\
       \affaddr{University of Illinois}\\
       \email{caesar@cs.illinois.edu}
\alignauthor Nikita Borisov\\
       \affaddr{Dept. of ECE}\\
       \affaddr{University of Illinois}\\
       \email{nikita@illinois.edu}
}

\begin{document}

\maketitle
\vspace{-0.3in}

\newcommand{\paragraphb}[1]{\vspace{0.03in}\noindent{\bf #1} }
\newcommand{\paragraphe}[1]{\vspace{0.03in}\noindent{\em #1}}
\newcommand{\paragraphbe}[1]{\vspace{0.03in}\noindent{\bf \em #1} }

\newcommand{\prateek}{\textcolor{black}}
\newcommand{\matt}[1]{{\color{black}{#1}}}
\newcommand{\nikita}{\textcolor{black}}
\newcommand{\mattc}[1]{}

\newcommand\mn[1]{\marginpar{\tiny \raggedright #1}}
\marginparwidth 36pt

\hyphenation{PlanetLab}
\begin{abstract}
Distributed hash tables suffer from several security and privacy vulnerabilities, 
including the problem of Sybil attacks. 
Existing social network-based solutions 
to mitigate the Sybil attacks in DHT routing have a high state requirement 
and do not provide an adequate level of privacy. For instance, such techniques require a 
user to reveal their social network contacts. 
We design X-Vine, a protection mechanism for distributed hash tables 
that operates entirely by communicating over social 
network links. 
As with traditional peer-to-peer systems, X-Vine provides 
robustness, scalability, and a platform for innovation.  The use of social 
network links for communication helps protect participant privacy and adds a new
dimension of trust absent from previous designs. X-Vine is resilient to denial of 
service via Sybil attacks, and in fact is the first Sybil defense that requires only 
a logarithmic amount of state per node, making it suitable for large-scale 
and dynamic settings. X-Vine also helps protect the privacy of users social 
network contacts and keeps their IP addresses hidden from those outside
of their social circle, providing a basis for pseudonymous communication. 
We first evaluate our design with analysis and simulations, using several
real world large-scale social networking topologies. We show that the 
constraints of X-Vine allow the insertion of only a logarithmic number of 
Sybil identities per attack edge; we show this mitigates the impact of malicious 
attacks while not affecting the performance of honest nodes. Moreover, our 
algorithms are efficient, maintain low stretch, and avoid hot spots in the 
network. 
\matt{We validate our design with a PlanetLab implementation and a Facebook
plugin.}  %
\end{abstract}

\section{Introduction}

Peer-to-peer (P2P) networks have, in a short time, revolutionized 
communication on the Internet.  One key feature of P2P networks 
is their ability to scale to millions of users without requiring 
any centralized infrastructure support. The best scalability and 
performance is offered by multi-hop distributed hash tables (DHTs), 
which offer a structured approach to organizing peers~\cite{chord,pastry,can,kademlia}.  
Multi-hop DHTs are the subject of much research and are also used in 
several mainstream systems~\cite{emule,vuze,coral}.

Securing DHTs has always been a challenging task~\cite{castro,dan-02,sit-02}, especially in the
face of a Sybil attack~\cite{sybil}, where one node can pretend to have multiple 
identities and thus interfere with routing operations. Traditional 
solutions to this attack require participants to obtain certificates~\cite{castro}, 
prove possession of a unique IP address~\cite{salsa,shadowwalker}, or perform some 
computation~\cite{borisov:p2p06}.  This creates a barrier to participation, limiting 
the growth of the P2P user base, and at the same time does not fully address the problem of Sybil 
attacks.

\matt{
To address this, recent research proposes to use social network trust
relationships to mitigate Sybil attacks~\cite{sybilguard,sybillimit,
sybilinfer}.  However, these systems share some key shortcomings:

\paragraphe{High control overhead:} These systems rely on flooding or large
numbers of repeated lookups to maintain state.  For example,
Whanau~\cite{whanau:nsdi10} is the state-of-art design that secures routing in
DHTs, but it is built upon a \emph{one-hop} DHT routing mechanism, and has high
overheads: state and control overhead increases with $O(\sqrt{n} \log n)$,
where $n$ is the number of participants in the social network.  As networked
systems become increasingly deployed at scale (e.g., in the wide area, across
service providers), in high-churn environments (e.g., developing regions,
wireless, mobile social networks~\cite{mobile-osn}), and for
applications with stronger demands on correctness and availability (e.g.,
online storage, content voting, reputation systems) the problem of high
overhead in existing works stands to become increasingly serious; multi-hop DHT
routing mechanisms are going to be necessary. 

\paragraphe{Lack of privacy:} These systems require a user to reveal social
contact information (friend lists).  Some of these schemes require global
distribution of this contact information. This is unfortunate, as social
contacts are considered to be private information: leading real-world systems
like Facebook~\cite{facebook} and LiveJournal~\cite{livejournal} provide users
with a functionality to limit access to this information.  Forcing users to
reveal this private information could greatly hinder the adoption of these
technologies.
}            

A second privacy concern, common to both traditional DHTs and ones that use social networking information, is that users must communicate directly with random peers, 
revealing their IP addresses.  This provides an
opportunity for the attacker to perform traffic analysis and compromise user privacy~\cite{bauer:wifs09,liberatore:conext10}. 
\prateek{Prior work~\cite{mittal:ccs08,wang:ccs10}%
has 
demonstrated} that a colluding
adversary can associate a DHT lookup with its lookup initiator, and thus infer
the activities of a user. A \emph{pseudonymous} routing mechanism can defend
against such attacks, and would be especially beneficial for privacy sensitive
DHT applications~\cite{freenet,shadowwalker}.

To address these shortcomings, we propose X-Vine, a protection mechanism
for large-scale distributed systems that leverages social network trust
relationships.  X-Vine has several unique properties. X-Vine protects {\em
privacy} of social relationships, by ensuring that a user's relationship
information is revealed only to the user's immediate friends. At the same time,
X-Vine also protects {\em correctness} of DHT routing, by mitigating Sybil
attacks while requiring only logarithmic state and control overhead.
\prateek{To the best of our knowledge, X-Vine is the first system to provide
both properties}, which may serve to make it a useful building block in
constructing the next generation of social network based distributed systems. Finally, X-Vine also provides a basis for pseudonymous
communication; a user's IP address is revealed only to his/her trusted social
network contacts. 

X-Vine achieves these properties by incorporating social network trust 
relationships in the DHT design.
Unlike traditional 
DHTs, which route directly between overlay participants (e.g.,~\cite{whanau:nsdi10}), 
X-Vine embeds the DHT directly into the social fabric, allowing communication through the DHT to
leverage trust relationships implied by social network links. This is done by using mechanisms 
similar to network layer DHTs like VRR~\cite{vrr}. We leverage this
structure for two purposes.  First, communication in X-Vine is carried out
entirely across social-network links.%
The use of social network links %
enables pseudonymous 
communication; while the recipient may know the opaque identifier (pseudonym) for the source, the IP address of the 
source is revealed only to his/her friends. Second, recent work has shown that social networks can be used to
detect Sybil attacks by identifying a bottleneck cut that connects the Sybil
identities to the rest of the network~\cite{sybilguard,sybillimit,sybilinfer}.
X-Vine enables comparable Sybil resilience by bounding the number of DHT
relationships that can traverse a particular edge.  With this multi-hop approach, we can
limit the number of Sybil identities per attack edge \prateek{(attack edges illustrated in Figure~\ref{fig:sybil-attack})} to logarithmic in the size
of the network with logarithmic control and routing state, a dramatic reduction
from previous Sybil defense approaches.  This allows X-Vine to scale to large
user bases and high-churn environments.

We evaluate X-Vine both analytically and experimentally using large scale 
real-world social network topologies. Since recent work~\cite{ravenben,vishwanath-wosn09} 
has advocated the use of interaction networks as a more secure realization of social 
trust, we also demonstrate the performance of X-Vine on interaction graphs. From our 
evaluation, we find that X-Vine is able to route using 10--15 hops (comparable to other DHTs) in topologies
with 100\,000 nodes while using only $O(\log n)$ routing state. In particular, 
we show that the overhead of X-Vine is two orders of magnitude smaller than Whanau.
With respect to Sybil resistance, we found that honest nodes are able to securely route to each other
with probability greater than $0.98$ as long as the number of attack edges is $g \in o(n/(\log n))$.  
Using an implementation on PlanetLab, we estimate the median lookup latency in a 100\,000 node 
topology to be less than 1.2 seconds. Even when 20\% of the nodes fail simultaneously, the lookups 
still succeed with a probability greater than 95\%. Finally, we also implement a plugin for DHT designers that can 
enable them to easily integrate social network contacts with a DHT by leveraging existing online social 
networks like Facebook.  

Our proposed techniques can be applied in a wide variety of scenarios that leverage DHTs:
\begin{ndss}
1.) Large scale P2P networks like Vuze/Kad/Overnet are popularly used for file sharing and content distribution. However, 
these networks are vulnerable to attacks on the routing protocol~\cite{attacking-kad} and do not protect the privacy 
of the user~\cite{mittal:ccs08}. X-Vine protects against 
attacks \prateek{that target} the DHT mechanisms and provides a basis for pseudonymous communication. 
Moreover, X-Vine is also robust to the high churn prevalent in these networks. 
2.) Applications like Coral~\cite{coral}, Adeona~\cite{adeona}, and Vanish~\cite{vanish} are built on top of DHTs. 
The security properties of these applications can often be compromised by exploiting vulnerabilities in the DHT. As an example, the security of 
Vanish was recently compromised by a low-cost Sybil attack on the Vuze network~\cite{defeating-vanish}. 
Our proposed techniques protect these applications by bounding the number of Sybil identities in the DHT. 
3.) Decentralized P2P anonymous communication systems like Tarzan~\cite{tarzan}, 
Salsa~\cite{salsa} and ShadowWalker~\cite{shadowwalker} assume an external Sybil defense 
mechanism. X-Vine is particularly suitable for designing Sybil-resilient P2P anonymous communication 
systems, since it provides secure as well as pseudonymous routing.  
4.) Freenet~\cite{freenet} is a widely used censorship resistant overlay network, but its routing algorithm has been shown 
to be extremely vulnerable in presence of even a few malicious nodes~\cite{pitch-black}. X-Vine can enable 
peers to resist censorship by securely and pseudonymously retrieving data objects from the Freenet network. 
5.) Membership concealing overlay networks (MCONs)~\cite{membership-concealing} hide the identities of 
the peers participating in a network (different from pseudonymity). Our proposed techniques can provide a substrate for designing fully 
decentralized membership concealing networks.
\end{ndss}   
 
\begin{techreport}
\begin{itemize}
\item Large scale P2P networks like Vuze/Kad/Overnet are popularly used for file sharing and content distribution. However, 
these networks are vulnerable to attacks on the routing protocol~\cite{attacking-kad} and do not protect the privacy 
of the user~\cite{mittal:ccs08}. X-Vine protects against 
attacks \prateek{that target} the DHT mechanisms and provides a basis for pseudonymous communication. 
Moreover, X-Vine is also robust to the high churn prevalent in these networks. 
\item Applications like Coral~\cite{coral}, Adeona~\cite{adeona}, and Vanish~\cite{vanish} are built on top of DHTs. 
The security properties of these applications can often be compromised by exploiting vulnerabilities in the DHT. As an example, the security of 
Vanish was recently compromised by a low-cost Sybil attack on the Vuze network~\cite{defeating-vanish}. 
Our proposed techniques protect these applications by bounding the number of Sybil identities in the DHT. 
\item Decentralized P2P anonymous communication systems like Tarzan~\cite{tarzan}, 
Salsa~\cite{salsa} and ShadowWalker~\cite{shadowwalker} assume an external Sybil defense 
mechanism. X-Vine is particularly suitable for designing Sybil-resilient P2P anonymous communication 
systems, since it provides secure as well as pseudonymous routing.  
\item Freenet~\cite{freenet} is a widely used censorship resistant overlay network, but its routing algorithm has been shown 
to be extremely vulnerable in presence of even a few malicious nodes~\cite{pitch-black}. X-Vine can enable 
peers to resist censorship by securely and pseudonymously retrieving data objects from the Freenet network. 
\item Membership concealing overlay networks (MCONs)~\cite{membership-concealing} hide the identities of 
the peers participating in a network (different from pseudonymity). Our proposed techniques can provide a substrate for designing fully 
decentralized membership concealing networks.
\end{itemize}
\end{techreport}

\begin{techreport}
\paragraphb{Roadmap:} The rest of the paper describes and evaluates 
X-Vine.  We start by giving a high-level overview of the problem we address and our
approach (Section~\ref{sec:overview}), followed by a detailed description of  
our routing algorithm (Section~\ref{sec:socdht}) and its security 
mechanisms (Section~\ref{sec:sybil-proof-dht}). We then describe our experimental 
results (Section~\ref{sec:analysis}). Finally, we summarize related work 
(Section~\ref{sec:related}), discuss X-Vine's limitations 
(Section~\ref{sec:limitations}), and conclude (Section~\ref{sec:conclusion}).
\end{techreport}

\section{X-Vine Overview}
\label{sec:overview}

\subsection{Design Goals}

\begin{ndss}
Our design has several key goals:
(i) {\em Secure routing:} if an honest node $X$ performs
a lookup for an identifier $ID$, then the lookup mechanism must
return the global successor of $ID$ (present in the routing tables
of honest nodes).
(ii) {\em Pseudonymous communication:} an adversary should not 
be able to determine the IP address corresponding to a user.
(iii) {\em Privacy of user relationships:} an adversary 
should not be able to infer a user's social contacts.  
(iv) {\em Low control overhead:} the control overhead of the system 
should be small to enable a scalable design.  This excludes flooding-based and
single-hop mechanisms. 
(v) {\em Fully decentralized design:} we target a fully decentralized architecture 
without any central points of trust/failure. 
We note that requirements (ii), (iii) and (iv) distinguish us from prior work---state-of-the-art approaches 
do not provide pseudonymous routing, do not preserve privacy of user relationships, and have 
high control overhead.  
\end{ndss}

\begin{techreport}
We start by defining the goals for our design.

\paragraphe{1. Secure routing:} if an honest node $X$ performs 
a lookup for an identifier $ID$, then the lookup mechanism must 
return the global successor of $ID$ (present in the routing tables 
of honest nodes).\\ 
\paragraphe{2. Pseudonymous communication:} an adversary should not 
be able to determine the IP address corresponding to a user.\\
\paragraphe{3. Privacy of user relationships:} an adversary 
should not be able to infer a user's social contacts.  \\
\paragraphe{4. Low control overhead:} the control overhead of the system 
should be small to enable a scalable design.  This excludes flooding-based and
single-hop mechanisms.                                 \\
\paragraphe{5. Low latency:} the length of the path used to 
route to an arbitrary identifier should be small, in order to 
minimize lookup latency.                       \\
\paragraphe{6. Churn resilience:} even when a significant fraction of nodes 
fail simultaneously, lookup queries should still succeed. \\
\paragraphe{7. Fully decentralized design:} we target a fully decentralized architecture 
without any central points of trust/failure. 

We note that requirements 2, 3 and 4 distinguish us from prior work---state-of-the-art approaches 
do not provide pseudonymous routing, do not preserve privacy of user relationships, and have 
high control overhead.  
\end{techreport}

\subsection{Threat Model and Assumptions}

We assume that a fraction of real users are compromised and colluding. Recent
work~\cite{sybilguard,sybillimit,sybilinfer} has leveraged the insight that it
is costly for an attacker to establish many trust relationships. Following this
reasoning, we assume that the number of attack edges, denoted by $g$, is
bounded. \prateek{Similar to prior work, we assume that the attack edges are not specially chosen.} 
We also assume that the set of colluding compromised nodes is a
Byzantine adversary, and can deviate from the protocol in arbitrary ways by
launching active attacks on the routing protocol. In particular, the set of
compromised nodes can launch a Sybil attack by trying to insert multiple fake
identities in the system.
\prateek{The key assumption we make about the adversary is that 
Sybil identities are distributed randomly in the DHT identifier space. 
We note that this assumption is a limitation of the X-Vine protocol, 
as discussed in Section~\ref{sec:limitations}. An exploration of defenses 
against adversaries who concentrate their nodes in a particular region of 
the overlay is beyond the scope of this paper.}

\subsection{Solution Overview}

\begin{figure}[!t]
\centering
\includegraphics[width=2.0in]{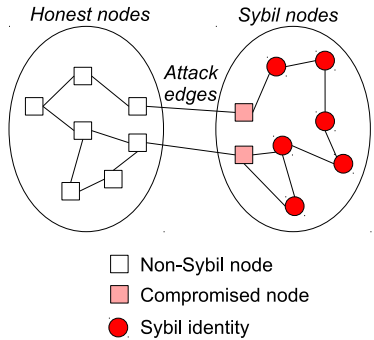}
\vspace{-0.1in}
\caption{Illustration of honest nodes, Sybil nodes, and attack edges between them.}
\label{fig:sybil-attack}
\vspace{-0.1in}
\end{figure}

We start by describing our algorithm in the context of an abstract, static
network.  Suppose we have a graph $G$, where
nodes correspond to users of the social network, and edges correspond to social
relationships between them.  Our approach runs a DHT-based routing scheme over
the graph that embeds path information in the network.  We first describe how
routing is performed, and then describe how the path information it creates can
be used to mitigate Sybil attackers.

\paragraphb{Pseudonymous routing in the social network:} We construct a DHT on top of the
social network, using mechanisms similar to network layer DHTs~\cite{vrr}. 
 Each node in the network selects a random numeric
identifier, and maintains paths ({\em trails}) to its neighbors in the
identifier space in a DHT-like fashion. To join the network, a node performs a
discovery process to determine a path to its {\em successors} in the DHT. Then,
the node embeds trails in the network that point back to the joining node's
identifier. To route messages, packets are forwarded along these trails. By
maintaining trails to each of the node's successors, a node can forward a
message to any point in the namespace. 
\matt{Users
that are directly connected by a social network link simply communicate via the
IP layer.}
All communication performed by a
node is done only with its friends, and this communication is done in a manner
that does not reveal the node's local topology, preventing leakage of
friendship list information to non-friends. Routing over social 
links also enables a user to communicate pseudonymous with respect to non-friends.

\paragraphb{Protecting against Sybils:} The scheme described above does not
mitigate the Sybil attack, as a malicious node can repeatedly join with different
identifiers, and thereby ``take over'' a large portion of the identifier space.
Malicious nodes can in fact pretend that there is an entire network of Sybil nodes
behind themselves (Figure~\ref{fig:sybil-attack}).  To protect against the Sybil
attack, we constrain the number of paths between honest nodes and malicious nodes. Since Sybil nodes by their
very nature are likely to be behind a small ``cut'' in the graph, by
constraining the number of paths that may be set up, we can constrain the
impact that a Sybil node can have on the entire network.  In particular,
honest nodes rate-limit the number of paths that are allowed to be constructed
over their adjacent links, thereby limiting the ability of Sybil nodes to join
the routing scheme, and hence participate in the network.  When a joining node
attempts to establish a trail over an edge that has reached its limit, the node
adjacent to the full link sends the joining node a message indicating failure
of the join request. This limits Sybil nodes from constructing very many paths
into the network.  Since Sybil nodes cannot construct many trails, they cannot
place many identifiers into the DHT. Hence, an honest node can send traffic to
another honest node by forwarding traffic over the DHT, as trails are unlikely
to traverse Sybil-generated regions of the network.

\section{X-Vine Protocol}
\label{sec:socdht}

The key feature of our design is that all DHT communication happens over social
network links.\footnote{Applications such as Vuze may optionally choose to
benefit only from X-Vine's Sybil resilience, and can forgo pseudonymity by
directly transferring files between overlay nodes after the lookup operation.}
By ensuring that all communication takes place over social network links, we
can leverage the trust relationships in the social network topology to enforce
security properties. A trivial security property of our design is that an
adversary needs to be connected to honest users via a series of social network
links to communicate with them. Moreover, the IP address of the nodes only
needs to be revealed to their contacts, enhancing privacy for users. Most
importantly, our design is able to effectively resist Sybil attacks even when
the number of attack edges is large.

\subsection{Routing Over Social Networks}
\label{sec:routing}

\begin{figure}[!t]
\centering
\includegraphics[width=3.5in]{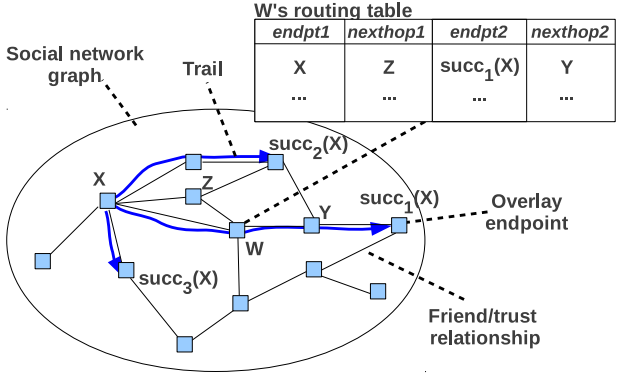}
\caption{Overview of X-Vine.}
\label{fig:grapevine}
\vspace{-0.2in}
\end{figure}

Figure~\ref{fig:grapevine} illustrates the design of X-Vine.
Our design uses a VRR-like~\cite{vrr} protocol to construct and maintain
state at the overlay layer.
Here, we first describe the state maintained by each node, and then describe how that state
is constructed and maintained over time. 

\paragraphb{State maintained by each node:} X-Vine constructs a {\em
social overlay} on top of the social network, where a node has direct links to
friends, but also maintains ``overlay links'' to remote nodes. These remote
nodes ({\em overlay endpoints}) are selected in a manner similar to
Chord~\cite{chord}: each node is assigned an identifier from a ring namespace,
and forms overlay links to nodes that are successors (immediately adjacent in
the namespace), and (optionally) fingers (spaced exponentially around the ring).  Unlike
Chord however, a node is not allowed to directly send packets to its overlay
neighbor: for security reasons, nodes only receive packets from their social
network links. Hence, to forward a packet from a node to one of its overlay
endpoints, the packet will have to traverse a {\em path} in the social network
graph.  To achieve this, corresponding to each of its overlay endpoints, a node
maintains a {\em trail} through the social network. Each node along the trail
locally stores a {\em record} consisting of four fields: the identifiers of the
two endpoints of the trail, and the IP addresses of the next and previous hops
along the trail.  Using this information, a node can send a packet to its
endpoints, by handing the packet off to the first node along the trail, which
looks up the next hop along the trail using its trail records, and so on.
Furthermore, using a Chord-like routing algorithm, a node can route to any
other node in the namespace, by (upon reaching an endpoint) selecting the next
overlay hop that maximizes namespace progress to the destination (without
overshooting).  As an optimization, instead of waiting until the endpoint is
reached to determine the next overlay hop, intermediate nodes along the path
may ``\emph{shortcut}'' by scanning all their trail records, and choosing the endpoint
that maximizes progress in the namespace (see Algorithm 1 in Appendix~\ref{sec:pseudocode}). If the intermediate node discovers an
endpoint that makes more namespace progress to the destination than the current
next overlay hop, the intermediate node may choose to forward the packet
towards this new endpoint, to speed its progress (while
explicitly maintaining the next overlay hop in the packet is not strictly necessary for routing, we do so to simplify parts of our design described later).

\paragraphb{State construction and maintenance:} Since nodes can route, we can
perform other DHT operations by simply performing routing atop this structure.
For example, we can execute a Chord-like join: upon arriving at the network, a
node can route a {\em join request} towards its own identifier, and the node
that receives it can return back the identifiers which should be the joining node's
successors.  However, there are two key changes we need to make.
First, when a node initially arrives, it does not yet have any trail state and
hence cannot forward packets. To address this, the joining node randomly selects one of
its friends in the social network to act as a {\em bootstrap} node.  The
joining node sends its join request using the bootstrap node as a proxy.
Second, the joining node also needs to build trails to each of its endpoints
(e.g., its successors). To do this, for each endpoint, it sends a {\em trail
construction request} to the identifier of that endpoint. As the request is
routed, each intermediate node along the path locally stores a record
corresponding to the trail.  Finally, when these steps are completed, the
joining node can route to any node in the network (by forwarding packets
through its endpoints), and it can receive packets from any node in the network
(by accepting packets through its endpoints).  To maintain this state, we need
to achieve two things. First, we would like to correctly maintain the set of
records along each trail in the presence of churn, so each node can reach the trail endpoint. This is
done in a manner similar to AODV~\cite{aodv}: each node along the path locally probes its
neighbors and removes trail records (sending {\em teardown} messages upstream
if necessary) corresponding to failed trails.  Second, we would like to make
sure each trail points to the corresponding globally correct successor/finger.
To do this, we leverage the stabilization mechanisms from Chord and
VRR~\cite{chord,vrr}.

\subsection{Balancing Routing State}
\label{sec:loadbalancing}

\paragraphb{Temporal correlation:} while the scheme above is correct, it performs poorly in practice. The reason
for this is due to {\em temporal correlation}---since trails are constructed using
other trails, social network links that are initially chosen to be part of a
trail become increasingly likely to be part of later trails. Because of this,
nodes that join the network early tend to accumulate more state over time.  To
illustrate this problem, we describe an example.  Suppose a node $X$ has $d$  
friends $a_1,a_2,..,a_d$. Suppose also that there is a trail from $X$
to $Y$ for which the next hop is node $a_d$. Next, suppose node $X$ is an
intermediate node in a new overlay path  that is being setup from node $a_1$
(which is also the previous hop). With probability $2/d$, the next hop of the
overlay path will be $a_d$. \matt{Similarly}, in the future, the
probability of $a_d$ being chosen as the next hop in an overlay path increases
to $3/(d+1)$, and then to $4/(d+2)$, and so on. This example illustrates
that a social network link that was initially chosen as part of a trail
has an increasing chance of being chosen in trails that are  set up in
the future. Consequently nodes that join the social network early tend to be
part of many trails. This is not desirable from both a
security perspective or a performance perspective.

\paragraphb{Stabilization algorithms:} To address the problem of temporal correlation, we propose 
\matt{two modifications to the core X-Vine algorithms: }
The first algorithm  leverages the social connections of new users to reduce the path
lengths of existing trails. When a new node joins the system, its social contacts that are already
part of the X-Vine system consider all trails in their routing tables that have a path length greater
than a threshold $\mathit{thr}_1$ (set to the upper quartile of trail path path lengths). Corresponding to each 
such trail, the social contacts check if modifying the trail via the new
node would reduce the path length, and if so, a teardown message is sent to the old trail and another trail via
the new node is setup. The threshold on the path length helps to avoid needless communication for trails
 that are already short, and are thus unlikely to benefit much from new edges in the social graph topology. 
The second algorithm helps to load balance the routing state at nodes, and also leads to a reduction in the
 path lengths of trails. This algorithm is run by all nodes whose routing state is greater than
a threshold $\mathit{thr}_2$. Such nodes consider all trails in their routing tables whose path length
is greater than a threshold $\mathit{thr}_1$ (similar to the previous algorithm), and send messages to the overlay end points to check
if alternate trails can be established, and if their path length is shorter than the current path length. If a shorter alternate
trail exists, then it replaces the existing trail. This helps reduce the routing state size at congested nodes,
while simultaneously reducing the trail path lengths.

\subsection{Bounding State With Local Policies}
\label{sec:bounding}
\begin{figure}
\centering
\fbox{
\includegraphics[width=2.1in]{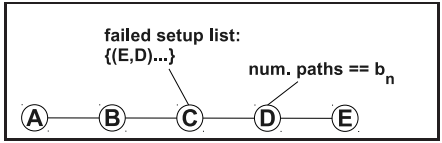}
}
\caption{Example: backtracking.}
\label{fig:backtrackeg}
\end{figure}

We have seen that the shortcut-based routing protocol described in Section~\ref{sec:routing} faces the problem of temporal correlation, leading to unbounded growth
in routing state. 
To complement our stabilization algorithms, we propose a mechanism by which nodes can set a hard bound on 
their routing state size using local routing policies. 
These policies can be set to account for heterogeneity across nodes, users' desired degree of participation in the network, and to limit the worst-case state overhead
at any particular node.
Our architecture allows users to 
set two types of policies pertaining to state maintained at their local node:%
{\em Bounding routes per link:} If the number of trails traversing an adjacent social network 
link reaches a threshold $b_l$, then the user's node refuses to set up 
any more trails traversing that link. 
{\em Bounding routes per node:} If the number of trails traversing the user's node reaches a threshold value $b_n$, 
then the node refuses to set up any more
trails via itself.  
Due to these routing policies, it is possible that a request to set up a trail 
 may be unable to make forward progress in the overlay namespace.
To address this, we introduce a technique called {\it backtracking} that
explores alternate social network paths in case an intermediate node refuses to
process a path setup request.  To do this, each node maintains a {\em failed
setup list}, containing a list of trails that have failed to set up.  When a
node attempts to set up a trail via a next hop, and receives a {\em rejection}
message indicating that the next hop is full, the node inserts an entry into
its failed setup list.  Each record in the list contains the identifier of the
destination overlay endpoint that the packet was traversing towards, and the identifier of
the next hop in the social network that rejected the path setup.  When
forwarding a message to a particular destination endpoint, a node removes from
consideration next hops contained in the failed setup list corresponding to
that endpoint (see Algorithm 2 in Appendix~\ref{sec:pseudocode}). The failed setup list is periodically garbage collected
by discarding entries after a timeout.

For example (Figure~\ref{fig:backtrackeg}), suppose node $A$ wishes to establish a
path to $E$, and determines $B$ is the best next overlay hop. 
$A$ places $E$ into the {\em next overlay hop} field in the
message, and forwards the message to $B$. Similarly, $B$ forwards
the message to $C$. Suppose $D$ is congested (has more than $b_n$ paths
traversing it). In this case, $C$ sends the path setup message to $D$, but $D$
responds back with a rejection message. $C$ then stores the entry $(E,D)$ in
its failed setup list, to indicate that establishing a path via $D$ to reach
$E$ was unsuccessful. $C$ then attempts to select an alternate next hop that
makes progress in the namespace (either a route to the current next overlay
hop, or a ``shortcut'' route that makes more progress than the current next
overlay hop). If $C$ does not have such a route, it sends a rejection message
back to $B$, which inserts the entry $(E,C)$ in its failed setup list. This
process repeats until a path is discovered, or a time-to-live (TTL) contained
in the packet is exceeded. When the TTL is exceeded, the path setup fails, and
the source node must attempt to rejoin to establish the path.

\section{Securing X-Vine}
\label{sec:sybil-proof-dht}

The previous section described our approach to perform routing atop the social
network.  In this section, we describe how to extend and tune the design in the
previous section to improve its resilience to attack.
We start by providing an overview of attacks on our design(Section~\ref{sec:attacks}),
and then propose extensions to improve resilience to them (Section~\ref{sec:defenses}).
\subsection{Attacks on the Routing Protocol}
\label{sec:attacks}
We investigate defenses to the following attacks on DHTs:

{\bf Sybil attack~\cite{sybil}:} The attacker can insert a large number of Sybil identities
in the DHT, and set up paths with their successors and predecessors. The attack
results in honest nodes' routing tables being populated by malicious Sybil
identities. This increases the probability that lookup queries will traverse
some malicious nodes, which can then drop or misroute the lookup queries.
Observe that to minimize resources, it suffices for Sybil identities to maintain
paths with only nodes in the predecessor list, since paths to the nodes in the
successor list will result in a shortcut to the honest successor nodes.  

{\bf Attacks on routing table maintenance:} In addition to the Sybil attack, 
the adversary could also manipulate the routing table
maintenance protocols to increase the probability of malicious nodes being present in 
honest nodes' routing tables. 
\paragraphe{Intercepting trails:} During churn, malicious nodes can become
part of a large number of trail paths between nodes, in order to attract 
lookup traffic (for example, by refusing to send trail teardown messages). 
\paragraphe{Attacking trail construction:} The attacker could prevent honest
nodes from finding a trail path to their correct successor. This could be
done by dropping or misrouting the trail setup messages. 
\paragraphe{Attacks on message integrity:} Malicious nodes that forward control traffic
could modify the content of the messages, to disrupt trail
setup (for example, by creating routing loops).  
\paragraphe{Forgery attacks:} The malicious nodes could spoof source identifiers in messages 
sent to honest nodes (for example, to give the appearance that the message came from the 
honest node's friends). 

{\bf Attacks on lookups:} Once the attacker is able to intercept a lookup query, 
it can either drop the 
packet or misroute it. Such attacks can prevent the honest nodes from either 
discovering their correct successor in the ring, or discovering a malicious 
successor list set respectively. By advertising malicious nodes as the successors 
of an honest joining node, a significant fraction of the honest joining node's 
traffic would traverse  malicious nodes. Note that attacks on both overlay 
construction and overlay routing are captured by this attack, since in a DHT, 
both bootstrap and routing are accomplished by the same operation: a lookup.    

\subsection{Proposed Defenses}
\label{sec:defenses}
We note that it is imperative to secure \emph{both} the routing table maintenance 
and 
lookup forwarding. If the routing table maintenance protocol
 were insecure, then the adversary could manipulate the routing table entries of honest 
nodes to point to malicious nodes, and routing to honest nodes would not be successful. 
However, even if the routing table maintenance mechanisms are secure, the adversary 
still has the opportunity to drop lookup packets or misroute them. 

{\bf Mitigating the Sybil attack:} To limit the impact of the Sybil attack, 
we propose that nodes implement a
routing policy that bounds the number of trails that traverse a social
network edge. We denote the bound parameter as $b_l$. Since the attacker has
limited attack edges, this bounds the number of overlay paths between the honest subgraph
and the Sybil subgraph {\em regardless of the attacker strategy}. Thus, we limit the 
number of Sybil identities that are part of the honest node's routing table. 
The key challenge in this approach is to determine the bound $b_l$ that enables most 
honest nodes to set up trails with each other while hindering the ability of Sybil nodes
to join the DHT. Our analytic and experimental results suggest that a bound of $b_l \in \Theta(\log n)$
works quite well. Similar to Yu et al.~\cite{sybillimit}, we assume that the bound $b_l$ is a system 
wide constant known to all honest nodes. 
Honest nodes are able to set up trails with each other even though there is a 
bound on the number of trails per social network link because of the fast-mixing nature of
the social network.  
On the other hand, a Sybil attack gives rise to a sparse cut in the social network topology, 
and we use this sparse cut to limit the impact of the Sybil identities.  
The number of overlay paths between the honest and Sybil subgraphs is bounded to $g \cdot b_l$. The adversary
could choose to allocate each overlay path to a different Sybil identity, resulting in $g \cdot b_l$ Sybil identities
in the DHT (in the routing tables of honest nodes). We can further limit the number of Sybil identities in the 
routing tables of honest nodes by ensuring that the adversary must allocate at least a threshold $t$ number of 
overlay paths per Sybil identity. This would bound the number of Sybil identities in honest nodes routing tables
to $g \cdot b_l / t$. Note that the number of overlay paths between the honest and Sybil regions does not change. 
\prateek{We propose the following mechanism to ensure that the adversary sets up trails with at 
least a threshold $t$ overlay neighbors. Nodes periodically probe their overlay neighbors to 
check if each successor in their routing table has set up a trail with at least $t$ other nodes in the overlay 
neighborhood. Note that the check is performed by directly querying the overlay neighbors. The threshold $t$ is 
set to  $t < 2 \cdot \mathit{num\_successors}$ to account for malicious overlay nodes returning incorrect replies. 
If the adversary does not allocate $t$ trails per Sybil identity (set up with its successors
and predecessors), the honest nodes can detect this via probing and can teardown the trails to the malicious 
Sybil identity. Note that the adversary cannot game the probing mechanism unless it has a large number of 
Sybil identities in the overlay neighborhood of a node. Since the Sybil identities 
are distributed at random in the overlay namespace, this is unlikely to happen unless the adversary has a 
large number of attack edges ($g \in \Omega(n/(\log n))$)}.

{\bf Securing routing table maintenance:} We provide the following defenses to attacks 
on routing table maintenance:

{\em Trail interception attacks:} Observe that our mechanism to defend against Sybil attacks, 
i.e., bounding the number of trails that traverse a social network link, also defends against malicious 
nodes that attempt to be a part of a large number of trails. Specifically, the adversary has a 
\emph{quota} of $g \cdot b_l$ trails between honest nodes and itself, and it can choose to utilize this 
quota either by inserting Sybil identities in the DHT or by being part of trails between two honest nodes. 
Either way, the effect of this attack is limited by the bound $b_l$.  

{\em Trail construction attacks:} Suppose that a node X is trying to set up a trail
with its overlay neighbor Y. To circumvent the scenario where a malicious intermediate node $M$ simply drops 
X's path set up request to Y, we propose that upon path setup the end point Y sends an acknowledgment along the 
reverse path back to X. If after a timeout, the node X does not receive an acknowledgment from Y, then it 
can retry sending the trail set up request over a different route. Again, the fast-mixing nature of the social
network topology guarantees that two nodes are very likely to have multiple paths between each other. 

{\em Message integrity and forgery attacks:} To provide message integrity is the use of 
self-certifying identifiers~\cite{self-certifying-filesystem,rofl,aip}. Nodes can append their public keys 
to the message and produce a digital signature of the message along with the appended public key. 
The self-certifying nature of identifiers ensures that the public key for a specified node identifier cannot be forged; 
this enables us to provide both message integrity as well as authentication.

{\bf Securing the lookup protocol:} Even if the routing table maintenance protocol is secure, 
the adversary can still drop 
\matt{or misroute lookup requests that traverse itself}.
We secure the lookup protocol using redundant routing, similar to Castro et al.~\cite{castro}. 
Instead of a single lookup, a node can choose to perform $r$ lookups for the destination (where $r$ is the 
redundancy parameter) using $r$ diverse trusted links in the social network topology.  Redundant routing increases the 
probability that at least one lookup will traverse only honest nodes and find the correct successor.  If the lookup 
is performed during route table maintenance, the correct successor can be identified since it will be impossible to
set up a trail to an incorrect one; if the lookup is searching for a particular node or data item, then self-certifying
techniques can be used to identify incorrect responses.

\subsection{Privacy Protection}

All communication in X-Vine happens over social network links; while a user's IP 
address is revealed to his/her social contacts, it is not exposed to random 
peers in the network. Therefore as long as a user's social contacts are trusted, 
he/she can communicate pseudonymously. %
Moreover, observe that X-Vine's mechanisms do not require a user to expose his/her social contacts. This is 
in sharp contrast to prior work~\cite{whanau:nsdi10}, wherein this information is revealed as part of protocol 
operations to everyone in the network. Note that in the absence of a mapping from a DHT ID to 
an IP address, the adversary cannot perform traffic analysis to infer social contacts. The only source 
of information leakage is when the adversary can map DHT IDs of two users to their respective 
IP addresses (for example, by virtue of being their trusted contacts); in this case the adversary 
can perform traffic analysis attacks to infer whether the two users have a trust relationship or not.  
In X-Vine, the privacy risk is with respect to social contacts, rather than random 
peers in the network.  Note that in this paper, we are only concerned with overlay level adversaries; adversaries 
which operate at the ISP level, or have external sources of information~\cite{narayanan:oakland09} are outside the scope of 
our threat model.

\section{Experiments and Analysis}
\label{sec:analysis}

We evaluate X-Vine with theoretical analysis, experiments
using real-world social network topologies, and \nikita{a prototype} implementation.
We measure routing state size, lookup path lengths,
security against Sybil attacks, resilience against churn, and lookup latency.
We also developed a Facebook application to facilitate the use of our design.

\paragraphb{Simulation environment:}
\matt{
We constructed an in-house event-driven simulator.
As done in~\cite{vrr}, we bootstrap X-Vine by selecting a random node in the social
network as the first node, and the social network neighbors of that
node then become candidates to join the X-Vine network.
Next, one of these neighbors is selected to join, with a probability
proportional to the number of trust
relationships it has with nodes that are already a part of the X-Vine network.
This process is then repeated.
Note that some nodes may not be successful in
joining because of the bound on number of trails per link (as
discussed in detail later).
}

\paragraphb{Data sets:}
Recent work has proposed the use of interaction graphs ~\cite{vishwanath-wosn09,ravenben} as a better indicator 
of real world trust than friendship graphs. 
Hence we consider both traditional social network graph topologies as well as interaction graph topologies in our evaluation. 
\begin{ndss}
We consider four datasets:
(i) {\em Facebook friendship graph from the New Orleans regional
network~\cite{vishwanath-wosn09}}, containing 
60\,290
nodes and 772\,843 edges. We processed the dataset in a manner similar to the
evaluation done in SybilLimit~\cite{sybillimit} and
SybilInfer~\cite{sybilinfer}, by imposing \prateek{a lower bound of $3$ and an upper 
bound of $100$ on the node degree (see ~\cite{sybillimit,sybilinfer} for details)~\footnote{Recent work by Mohaisen et al.~\cite{mohaisen:imc10} shows that 
social networks may not be as fast mixing as previously believed. However, we note that their results 
do not directly apply to X-Vine since they did not consider node degree bounds in their analysis. 
X-Vine excludes users having few friends from participating in the routing protocol, though such users could use their trusted friends to lookup keys.}}. 
(ii) {\em Facebook wall post interaction graph from the New Orleans regional
network~\cite{vishwanath-wosn09}}, containing 29\,140 users and 161\,969 edges
after processing. Note that links in this dataset are directed, and we consider an edge
between users only if there were interactions in both directions.
(iii) {\em Facebook interaction graph from a moderate-sized
regional network~\cite{ravenben}}.
The dataset consists of millions of nodes and edges; to 
account for scaling limitations of our experiments, we truncate 
the dataset by considering only a four hop
neighborhood from a seed node. After processing, we are left with 103\,840 nodes
and 961\,418 edges.
(iv) {\em Synthetic scale-free graphs:} Social networks exhibit a scale-free
node degree topology~\cite{scale-free}. Our network synthesis algorithm replicates   this
structure through preferential attachment~\cite{nagaraja}. %
The use of synthetic scale free topologies enables us
evaluate X-Vine while varying the number of nodes in the network.
\end{ndss}
\begin{techreport}
The datasets that we use have been summarized in Table ~\ref{table:datasets}.

{\em Facebook friendship graph from the New Orleans regional
network~\cite{vishwanath-wosn09}:} The original dataset consists of 60\,290
nodes and 772\,843 edges. 
We processed the dataset in a manner similar to the
evaluation done in SybilLimit~\cite{sybillimit} and
SybilInfer~\cite{sybilinfer}, by imposing \prateek{a lower bound of $3$ and an upper 
bound of $100$ on the node degree (see ~\cite{sybillimit,sybilinfer} for details)~\footnote{Recent work by Mohaisen et al.~\cite{mohaisen:imc10} shows that 
social networks may not be as fast mixing as previously believed. However, we note that their results 
do not directly apply to X-Vine since they did not consider node degree bounds in their analysis. 
X-Vine excludes users having few friends from participating in the routing protocol, though such users could use their trusted friends to lookup keys.}}. 
After processing, we are left with 50\,150 nodes and 661\,850 edges.

{\em Facebook wall post interaction graph from the New Orleans regional
network~\cite{vishwanath-wosn09}:} The original dataset consists of 60\,290
users. After processing, we are left with 29\,140 users and 161\,969 edges.
Note that links in this dataset are directed, and we consider an edge
between users only if there were interactions in both directions.  

{\em Facebook interaction graph from a moderate-sized
regional~\footnote{Because of privacy reasons, the name of the regional network
has been left anonymous by the authors of~\cite{ravenben}.}
network~\cite{ravenben}:} The dataset consists of millions of nodes and edges,
but our experiments are memory limited and do not scale to millions of nodes.
Instead, we first truncate the dataset by considering only a four hop
neighborhood from a seed node. After processing, we are left with 103\,840 nodes
and 961\,418 edges.  

{\em Synthetic scale-free graphs:} Social networks exhibit a scale-free
node degree topology~\cite{scale-free}. Our network synthesis algorithm replicates this
structure through preferential attachment, following the methodology of
Nagaraja~\cite{nagaraja}. The use of synthetic scale free topologies enables us
evaluate X-Vine while varying the number of nodes in the network. 

\begin{table}[!ht]
\caption{Topologies}
\label{table:datasets}
\centering
\small
\begin{tabular}{|p{3.25cm}|p{.9cm}|p{.9cm}|p{1.2cm}|}
\hline
 {\bf Dataset} & {\bf Nodes} & {\bf Edges} & {\bf Mean Degree} \\
\hline \hline
 {New Orleans Facebook Friendship graph}   & 50\,150 & 661\,850  & 26.39 \\ \hline
 {New Orleans Facebook Interaction graph } & 29\,140 & 161\,969 & 11.11 \\ \hline
 {Anonymous Facebook Interaction graph}  & 103\,840 & 961\,418 & 18.51 \\ \hline
\end{tabular}
\end{table}
\end{techreport}

\begin{ndss}
\begin{figure}[t]
\centering
\includegraphics[height=1.4in]{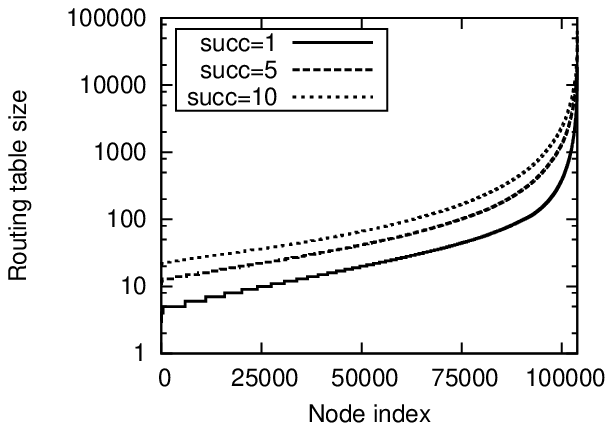}
\label{fig:state-nobounds}
\vspace{-0.1in}
\caption{{\em Routing state, with no bounds on state:} Anonymous Interaction graph.
}
\vspace{-0.1in}
\end{figure}
\begin{figure}
\centering
\includegraphics[height=1.4in]{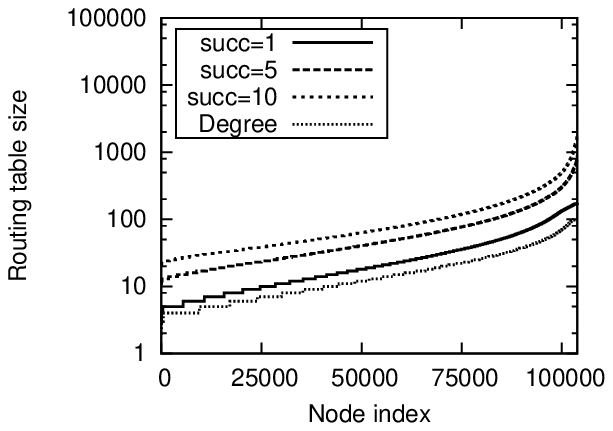}
\label{fig:state-bounds}
\vspace{-0.1in}
\caption{{\em Routing state, with node and edge bounds:}
Anonymous Interaction graph. (X-Vine's overhead is atleast two orders of magnitude less than Whanau). 
}
\vspace{-0.1in}
\end{figure}
\end{ndss}
\begin{techreport}
\begin{figure*}[t]
\centering
\mbox{
\hspace{-0.2in}
\hspace{-0.12in}
\begin{tabular}{c}
\psfig{figure=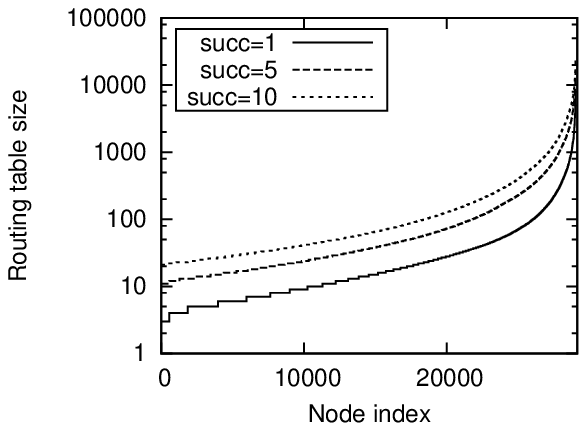,width=2.3in}\vspace{-0.00in}\\
{(a)}
\end{tabular}
\begin{tabular}{c}
\psfig{figure=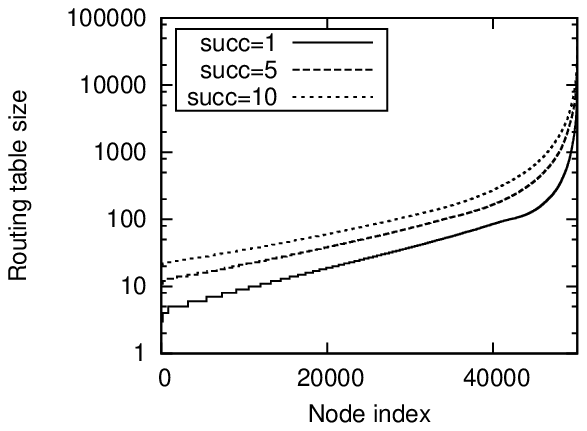,width=2.3in}\vspace{-0.00in}\\
{(b)}
\end{tabular}
\hspace{-0.12in}
\begin{tabular}{c}
\psfig{figure=figs/103840-nobounds.eps,width=2.3in}\vspace{-0.00in}\\
{(c)}
\end{tabular}
}
\caption{{\em Routing state, with no bounds on state:} (a) New Orleans Interaction graph, (b) New Orleans Friendship graph, and the (c) 
Anonymous Interaction graph.
Due to temporal correlation, some nodes exhibit high state requirements. }
\label{fig:state-nobounds}
\vspace{-0.13in}
\end{figure*}
\end{techreport}
\begin{techreport}
\begin{figure*}[t]
\centering
\mbox{
\hspace{-0.2in}
\hspace{-0.12in}
\begin{tabular}{c}
\psfig{figure=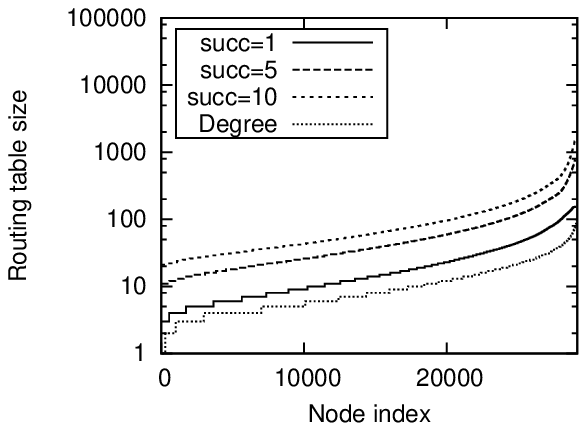,width=2.3in}\vspace{-0.0in}\\
{(a)}
\end{tabular}
\begin{tabular}{c}
\psfig{figure=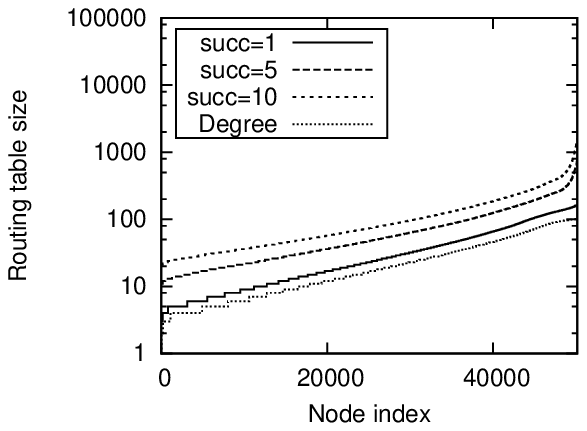,width=2.3in}\vspace{-0.0in}\\
{(b)}
\end{tabular}
\hspace{-0.12in}
\begin{tabular}{c}
\psfig{figure=figs/103840-bounds_logn.eps,width=2.3in}\vspace{-0.00in}\\
{(c)}
\end{tabular}
}
\caption{{\em Routing state, with node and edge bounds:}
(a) New Orleans Interaction graph, (b) New Orleans Friendship graph, and (c) Anonymous Interaction graph. Bounding state significantly reduces state requirements. Using a successor list of size 5, the average routing state for the three topologies is 67, 81, and 76 records respectively. 
X-Vine requires orders of magnitude less 
state than Whanau~\cite{whanau:nsdi10}.}
\label{fig:state-bounds}
\vspace{-0.1in}
\end{figure*}
\end{techreport}
\paragraphb{Overhead:}
Figure~\ref{fig:state-nobounds} plots the
routing table size for different successor list sizes.
We can see the temporal
correlation effect here, as the distribution of state \nikita{shows
super-exponential growth}.
\prateek{Temporal
correlation is highly undesirable both from a performance and a security
standpoint.} If the few nodes with very large state become unavailable due to
churn, the network could get destabilized. Moreover, if one of these nodes is
malicious, it could easily intercept a large number of lookups and drop them.
To address this, we enable the routing policy that bounds the number of paths traversing nodes and links. Based on our
analytic model in Appendix~\ref{sec:appendix}, we propose the following bound on
the number of paths per link: $b_l = \alpha \cdot  2 \cdot
\mathit{num\_successors} \cdot \log(n)$, where $\alpha$ is a small fixed
constant.  The bound per link ensures that if a node has degree $d$, then its
routing table size will never exceed $d \cdot b_l \in O(\log n)$. We can see
that the routing state does not undergo an           exponential increase as in
previous plots. Moreover, routing state increases with node degrees, which is
desirable.  \prateek{Based on these routing table sizes, we can estimate the
communication overhead of X-Vine by computing the cost of sending
heartbeat traffic for all records in the routing table.
Considering the routing table size to be $125$ records,
UDP ping size to be $40$ bytes, and a    heartbeat interval of $1$\,s, the
estimated mean communication overhead is only $4.8$\,KBps.}

\paragraphb{Comparison with Whanau~\cite{whanau:nsdi10}:} Routing state in
Whanau depends on the number of objects stored in the DHT. Routing tables in
Whanau are of size $\Theta(\sqrt n_o \log n_o)$, where $n_o$ is the number of
objects in the DHT. If there are too many objects stored in the DHT, Whanau
resorts to maintaining information about all the nodes and edges in the social
network \matt{(increasing state/overhead to $\Theta(n)$)}. 
If there are too few objects in the DHT, Whanau resorts to flooding to find
objects~\cite{whanau:nsdi10}. We note that such properties make Whanau
unsuitable for many common applications. Even if we consider the case where
each node in the DHT stores only tens of objects, the average routing table
size in Whanau for the $103\,840$ node anonymous interaction graph is about
$20\,000$ records---an \matt{increase of more than two orders of magnitude as
compared with X-Vine.} If we consider a heartbeat interval of 1 second in
Whanau (in order to accurately maintain object states for common DHT
applications), the resulting communication overhead is about 800\,KBps.  This
difference \matt{increases further} with an increase in the number of objects
in the DHT \matt{or the size of the network}.  For instance, we scaled up our
experiments to a larger $613\,164$ node anonymous interaction graph topology
using a machine with $128$\,GB RAM, and found that the average routing state in
X-Vine using a successor list size of $10$ was only $195$ records, as compared
with more than $50\,000$ records in Whanau. (Note that routing state in X-Vine
is independent of the number of objects in the DHT.)

\paragraphb{False Positive Analysis:} 
\prateek{Next, we consider the impact of link/node path bounds on honest node's 
ability to join the DHT. We found that most honest nodes were able to join the DHT 
due to the fast mixing nature of honest social networks.}
In fact, for all our experimental scenarios, the false-positive rate was less
than $0.5\%$, which is comparable to the state-of-the-art
systems~\cite{sybillimit,sybilinfer}.  By tuning the parameter $b_l$, it is
possible to trade off the false-positive rate for Sybil resilience: $b_l$ will
reduce the false-positive rate at the cost of increasing the number of Sybil
identities in the system. For the remainder of the paper, we shall use
$\alpha=1,\beta=5$.

\paragraphb{Path Length Analysis:}
\matt{
Table~\ref{table:pathlength} depicts the mean lookup path lengths for the real world datasets with varying successor list sizes and varying
redundancy parameter. We first observe that lookup performance improves with         increasing successor list sizes. For example, in the
New Orleans interaction graph, the mean lookup path length decreases from $97.9$ to  $15.4$ when the successor list size increases from
$1$ to $20$ (using $r=1$). Further improvements in performance can be realized by    performing redundant lookups as described in
Section~\ref{sec:sybil-proof-dht} and caching the lookup with the smallest path      length. We can see that in the same dataset, mean
lookup path length decreases from $15.4$ to $10.3$ when the redundancy parameter is  increased from $r=1$ to $r=5$ (using successor
list of size $20$). Further increases in redundancy show diminishing returns.
}
\begin{techreport}
Observe that when the successor list size is at least $10$, and the redundancy parameter is at least $10$, then the mean lookup path 
lengths for all datasets are less than  $15$ hops. Increasing the successor list size to $20$ (and keeping $r=10$) reduces this value to 
less than $11.5$ for all datasets.
\end{techreport} 

\begin{table*}[!t]
\caption{Mean Lookup Path Length}
\label{table:pathlength}
\centering
\small
\scriptsize
\begin{tabular}{|l|p{35pt}|p{35pt}|p{35pt}|p{35pt}|p{35pt}|p{35pt}|p{35pt}|p{35pt}|p{35pt}|p{35pt}|l}
\hline
 {\bf \# Succ} & \multicolumn{3}{|c|}{\bf New Orleans interaction graph} & \multicolumn{3}{|c|}{\bf New Orleans friendship graph} & \multicolumn{3}{|c|}{\bf Anonymous interaction graph} \\
	& {\bf $r=1$} & {\bf $r=5$ } & {\bf $r=10$}
	& {\bf $r=1$} & {\bf $r=5$ } & {\bf $r=10$}
	& {\bf $r=1$} & {\bf $r=5$ } & {\bf $r=10$} \\ \hline \hline
 {\bf 1}    & 97.9 & 57.7 & 51.7  & 103.6 & 57.5 & 48.1 & 166.7 & 96.3  & 81.0  \\ \hline
 {\bf 5}    & 30.0 & 18.2 & 16.8  & 34.8  & 19.3 & 16.7 & 48.9 & 25.5 & 21.7  \\ \hline
 {\bf 10}   & 20.2 & 13.0 & 12.16 & 23.1  & 13.7 & 12.1 & 29.9 & 16.9 & 14.8  \\ \hline
 {\bf 20}   & 15.4 & 10.3 & 9.6   & 17.0  & 10.7 & 9.45 & 21.0 & 12.8 & 11.3  \\ \hline
\end{tabular}
\end{table*}

\begin{figure*}[t]
\centering
\mbox{
\hspace{-0.2in}
\hspace{-0.12in}
\begin{tabular}{c}
\psfig{figure=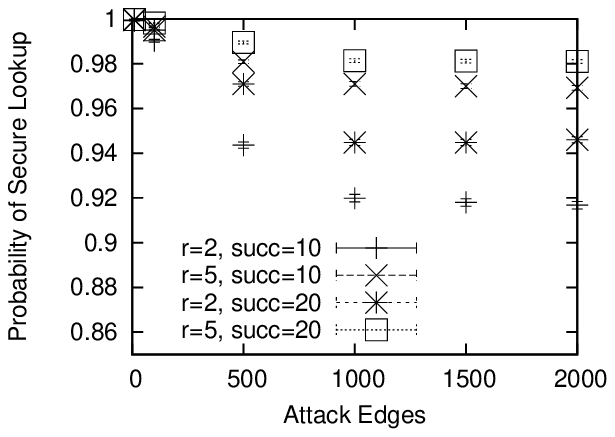,width=2.1in}\vspace{-0.00in}\\
{(a)}
\end{tabular}
\begin{tabular}{c}
\psfig{figure=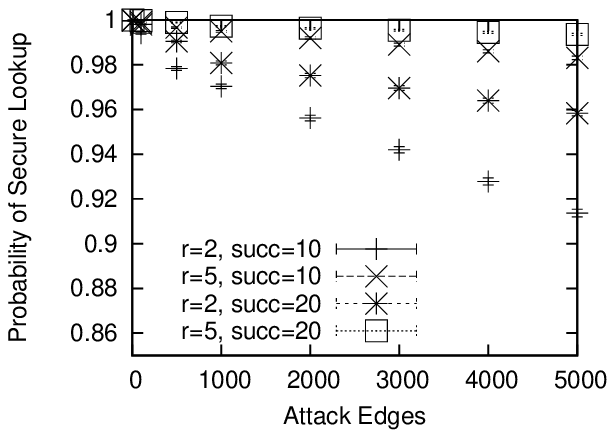,width=2.1in}\vspace{-0.00in}\\
{(b)}
\end{tabular}
\hspace{-0.12in}
\begin{tabular}{c}
\psfig{figure=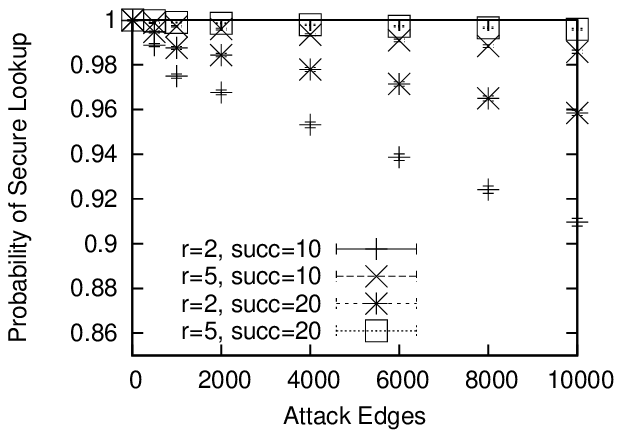,width=2.1in}\vspace{-0.00in}\\
{(c)}
\end{tabular}
}
\vspace{-0.1in}
\caption{Probability of secure lookup as a function of number of attack edges for (a) New Orleans interaction graph, (b) New Orleans friendship 
graph, and (c) Anonymous Interaction graph.}
\vspace{-0.2in}
\label{fig:attack}
\end{figure*}

\paragraphb{Security under Sybil Attack:} Recall
that if the adversary has $g$ attack edges, then the number of trails between 
the honest and the Sybil subgraph is bounded by $g \cdot b_l$ (regardless of the 
attacker strategy). Our attack methodology
is as follows: we randomly select a set of compromised nodes until the adversary 
has the desired number of attack edges. The compromised nodes then launch 
a Sybil attack, and set up trails between Sybil identities and their 
overlay neighbors. If the trail set up request starting from a Sybil node gets
shortcutted back to the Sybil \prateek{identities}, the request is backtracked. This ensures
that the adversary uses only a single attack edge per trail. 
Node identifiers of Sybil identities are chosen at random 
with the adversarial goal of intercepting as many lookups as possible. All lookups 
traversing compromised/Sybil nodes are considered unsuccessful. 

Figure~\ref{fig:attack} plots the probability of a secure lookup as a function of number of
attack edges, redundancy parameter, and size of
successor list. We find that the probability of secure lookup increases
as the redundancy parameter is increased. 
This is because as the number of
redundant lookups increases, there is a greater chance that a lookup will
traverse only honest nodes and return the correct result.  We also find that
the probability of secure lookup also increases when the size of the successor
list increases.  
This is because increasing
successor list size reduces the mean lookup path length, reducing the
probability that an adversary can intercept the lookup query. As long as
\nikita{$g \in o(n/(\log n))$}, the probability of secure lookup can be made
arbitrarily high by increasing the redundancy parameter and the successor list
size.  Finally, reducing $b_l$ would further limit the impact of Sybil
identities,  at the cost of increased false positives.

\begin{figure*}[t]
\centering
\mbox{
\hspace{-0.2in}
\hspace{-0.12in}
\begin{tabular}{c}
\psfig{figure=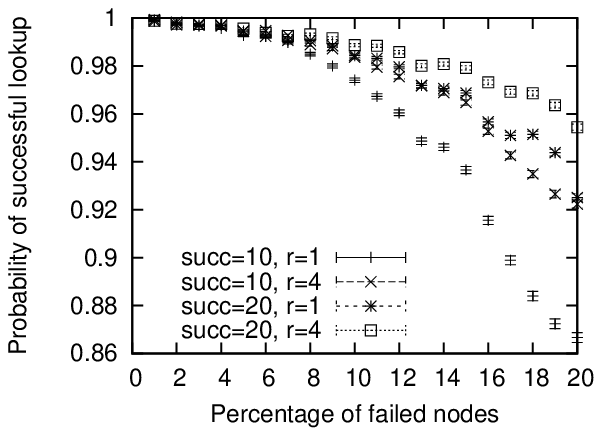,width=2.1in}\vspace{-0.0in}\\
{(a)}
\end{tabular}
\begin{tabular}{c}
\psfig{figure=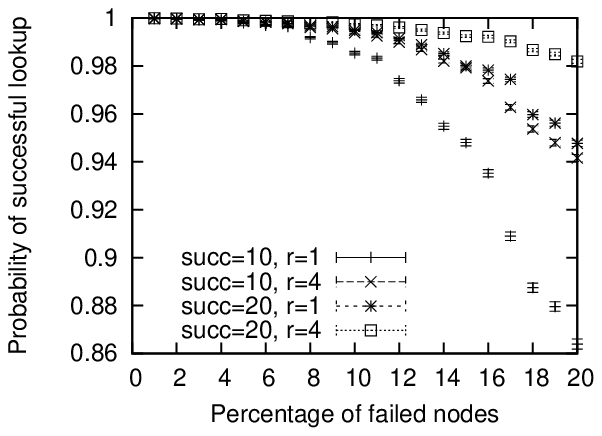,width=2.1in}\vspace{-0.0in}\\
{(b)}
\end{tabular}
\hspace{-0.12in}
\begin{tabular}{c}
\psfig{figure=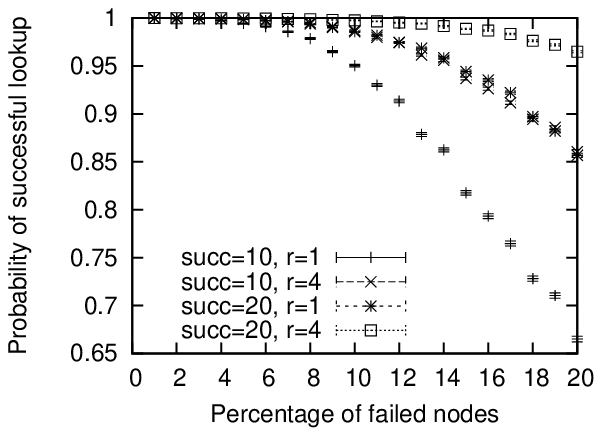,width=2.1in}\vspace{-0.00in}\\
{(c)}
\end{tabular}
}
\vspace{-0.1in}
\caption{{\em Lookup resilience against churn:}
(a) New Orleans Interaction graph, (b) New Orleans Friendship graph, and (c) Anonymous Interaction graph.}
\label{fig:churn}
\vspace{-0.2in}
\end{figure*}

\paragraphb{Churn Analysis:}
Next, we evaluate the performance of X-Vine under churn. We are interested in the \emph{static resilience} of 
X-Vine, i.e., the probability of lookup success after a fraction of the nodes in the system fail simultaneously. 
To account for churn, we modified the lookup algorithm to \emph{backtrack} whenever it cannot make forward progress in 
the overlay namespace. Figure~\ref{fig:churn} depicts the mean probability of lookup success as a function of the fraction 
of nodes that fail simultaneously, averaged over $100\,000$ lookups. Similar to the analysis of lookup security, we can 
see that an increase in either the redundancy parameter or the successor list size result in improved resilience against churn. 
We can also see that as the fraction of failed nodes increases, the probability of lookup success decreases, but is still greater 
than $0.95$ for all scenarios using $r=4$ and $\mathit{succ}=20$. 

\paragraphb{PlanetLab Implementation:}
To validate our design and evaluate lookup latency in real-world environments,
we implemented the X-Vine lookup protocol in C++ as a single-threaded program using 3\,000 LOC. 
We used libasync~\cite{libasync1,libasync2} and Tame~\cite{tame} to implement non-blocking socket functionality (UDP) in an event-based fashion.
\begin{figure}
\centering
\fbox{
\includegraphics[height=1.4in]{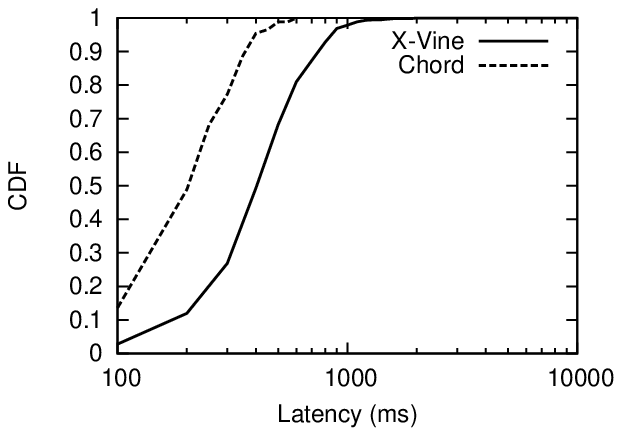}
}
\caption{Lookup latency}
\label{fig:latency}
\vspace{-0.1in}
\end{figure}
We ran our implementation over $100$ randomly selected nodes in the PlanetLab network. 
\prateek{We used a synthetic scale free graph as the social network topology.}  The duration 
of the experiment was set to 1 hour, and nodes performed lookups every 1 second. 
Figure~\ref{fig:latency} depicts the CDF of observed \prateek{one-way} lookup latencies. We can see that 
the median lookup latency was only $400$\,ms \prateek{(as compared to $200$\,ms in Chord)}, for the mean lookup path length of $5$ hops 
(not shown in the Figure). Using these values, we can estimate the median lookup latency 
for mean lookup path lengths of $10$ hops and $15$ hops (that were observed in our experiments 
over real world social network topologies in Table~\ref{table:pathlength}) to be 
about $800$\,ms and $1200$\,ms respectively. We see some outliers in Figure~\ref{fig:latency}
 due to the presence of a few slow/unresponsive nodes in PlanetLab. \prateek{For this experiment, 
we mapped vertices in the social network topology to random PlanetLab nodes (possibly in 
different geographic locations). Thus, our analysis is conservative; accounting 
for locality of social network contacts would likely improve the lookup performance.}

\paragraphb{Facebook Application:}
To bootstrap a X-Vine node, its user needs to input the IP addresses of his/her friends.
Since this can be a cumbersome for a user, we implemented a Facebook application (available at \url{http://apps.facebook.com/x--vine}) that 
automates this process and improves the usability of our design. The work flow of the application is as follows: 
(i) When a user visits the Facebook application URL, Facebook checks the 
credentials of the user, the user authorizes the application, and then the 
request gets redirected to the application hosting server.
(ii) The application server authenticates itself, and is then able to query 
Facebook for user information. The application server records the user information 
along with the user IP address.
(iii) The application server then queries Facebook for a list of user's friends, 
and returns their previously recorded IP addresses (if available) to the user.

This list of IP addresses could then be used by the DHT software to bootstrap 
its operations. Our implementation demonstrates that a user's social contacts 
can be integrated into the DHT protocol using only a few hundred lines of glue code. 
\prateek{Keeping in spirit with our fully decentralized design goal, in future,} 
our application could be implemented on a decentralized platform like 
Diaspora~\cite{diaspora} such that the app server is not a central point of trust or failure. 
\section{Related Work}
\label{sec:related}

X-Vine provides {\em multi-hop} social network routing, logarithmic state Sybil
defense, protects privacy %
of
friendship information, and
enables pseudonymous communication. %
Our work is the first to
provide these properties. 
However, X-Vine
builds upon foundational work in several areas:

\paragraphb{Sybil defense:}
\begin{ndss}
Sybil defenses must impose a cost on participation in the
network~\cite{sybil}, for example by requiring a central point of trust~\cite{castro},
or verifying resource expenditures such as CPU computation or possession of
a unique IP address~\cite{borisov:p2p06,rowaihy:infocom-mini07}.
All these solutions face a tradeoff between creating too high a barrier for
participation by honest users and making it too easy for malicious users to create   Sybil identities.
More recent work has recognized that it is expensive for %
an attacker
to establish trust relationships with honest users and thus
social network topologies can be used to detect and mitigate 
Sybil attacks.
SybilGuard~\cite{sybilguard} and SybilLimit~\cite{sybillimit}
use {\it random routes} for Sybil defense.
SybilInfer~\cite{sybilinfer} provides a Bayesian inference based algorithm for labeling nodes in a
social network as honest users or Sybils controlled by an adversary.
These systems are standalone Sybil defenses and do not provide a DHT functionality.%
\end{ndss}
\begin{techreport}
 Sybil defenses must fundamentally impose a cost on participation in the 
 network~\cite{sybil}.  One approach, advocated by Castro et al.~\cite{castro}, 
 requires users to provide identity credentials and/or payment to a 
 centralized authority, who then issues certificates allowing users to participate.  
 This authority, of course, becomes a central point of trust.  Decentralized approaches 
 instead allow nodes to directly verify some resource expenditure by other nodes, 
 such as CPU computation, or the possession of a unique IP address~\cite{borisov:p2p06,rowaihy:infocom-mini07}.  
 All these solutions face a tradeoff between creating too high a barrier for 
 participation by honest users and making it too easy for malicious users to create Sybil identities.
 More recent work has recognized that it is expensive for a malicious 
 adversary to establish trust relationships with honest users and thus 
 social network topologies can be used to detect and mitigate social 
 Sybil attacks.  The design of X-Vine is based upon the same principle.

SybilGuard~\cite{sybilguard} and SybilLimit~\cite{sybillimit} are 
decentralized systems for Sybil defense. 
These systems use special random walks called {\it random routes} for Sybil defense. 
In SybilLimit, as
long as the number of attack edges is less than a 
threshold ($g=o\left(\frac{n}{\log n}\right)$), then with high probability, 
a short random walk of $O(\log n)$ steps
is likely to stay within the set of honest nodes. 
Nodes in SybilLimit 
perform $\sqrt e$ short random walks (where $e$ is the number of edges 
 amongst the honest nodes) and keep track of their last edges (tails).  
By the birthday paradox, two honest nodes will share a common tail with high 
probability.  Each node allows only a certain number of random routes to traverse it, 
thereby limiting the number of Sybil identities that are validated by the honest nodes.

SybilInfer~\cite{sybilinfer} provides an algorithm for labeling nodes in a 
social network as honest users or Sybils controlled by an adversary. It 
takes as an input a social graph $G$, and generates a set of traces using short 
 random walks. Using a mathematical model of an honest social network, it performs 
 Bayesian inference to output a set of dishonest nodes. The Bayesian inference approach 
 can even be used to assign probabilities to nodes of being honest or dishonest. 
These systems are standalone Sybil defenses and do not provide a DHT functionality.%
\end{techreport}

Whanau~\cite{whanau:nsdi10} is the state of art Sybil resilient DHT~\cite{sybil-proof-dht,srdhtr} %
where nodes can communicate with only one 
intermediate hop. Each node performs $\sqrt e$ random walks to sample nodes for 
construction of their routing tables
\begin{techreport}; the Sybil resistant property of short 
random walks ensures that a high fraction of the sampled nodes are honest
\end{techreport}
. 
By querying routing table entries, nodes can construct their successor lists. As 
compared to X-Vine, 
Whanau provides its properties at the cost of maintaining $\sqrt n_o \log n_o$ state at each node (where $n_o$ 
is the number of objects).
The large state requirements mean that the system has difficulty maintaining accurate 
state in face of object churn. %
Whanau also requires 
the \matt{entire} social graph to be public, presenting significant privacy concerns. In contrast, 
X-Vine builds upon {\em network-layer} DHTs, embedding the DHT directly into the 
social network fabric. This enables X-Vine to provide good security while achieving 
improved scalability and privacy of social relationships. Moreover, X-Vine %
provides support for 
pseudonymous communication.   

\begin{techreport}
The concept of a bottleneck cut between a
fast-mixing social network and Sybil nodes has been used in a number of other
systems, such as SumUp~\cite{sumup-nsdi09}, a protocol for  online content
rating that is resilient to Sybil attacks; Ostra~\cite{ostra-nsdi08}, a system
to prevent unwanted communication from nodes; and
Kaleidoscope~\cite{kaleidoscope}, a system for censorship resistance.
\end{techreport}

\paragraphb{Security and privacy in DHTs:}
Other work deals with the issue of secure routing when a fraction 
of nodes in the DHT are 
\begin{techreport}
compromised~\cite{castro,sit-02,dan-02,salsa,halo}. 
Sit and Morris~\cite{sit-02}, as well as Wallach~\cite{dan-02} discuss security issues 
in DHTs. Castro~\cite{castro} proposed the use of redundant routing to improve the lookup 
 security. Nambiar and Wright~\cite{salsa} showed that redundant lookups in Chord may traverse 
 a few common nodes, and thus a few malicious nodes could subvert all of the redundant lookups.
They designed the Salsa DHT in which redundant lookups are unlikely to traverse common nodes. 
Kapadia and Triandopoulos~\cite{halo} propose to make redundant routes diverse by making use of 
 the observation that to perform a lookup for $A$, it suffices to lookup the nodes which have $A$ 
 as its finger, and then query them.
 Unlike X-Vine these
\end{techreport}
\begin{ndss}
compromised, for example by performing redundant lookups~\cite{castro,salsa,halo}.
Unlike X-Vine, these
\end{ndss}
systems are not concerned with the problem 
of Sybil attacks. 
Another line of research deals with the privacy of the DHT lookup. Borisov~\cite{borisov-thesis} as 
well as Ciaccio~\cite{ciaccio:pet06} proposed to incorporate anonymity into the lookup, but their algorithms 
do not consider active attacks. More recently, anonymous and secure lookups were considered 
in the designs of Salsa~\cite{salsa}, NISAN~\cite{nisan}, and Torsk~\cite{torsk}. However, 
recent work~\cite{mittal:ccs08,wang:ccs10} showed vulnerabilities 
in all the three designs. X-Vine improves the privacy of a user by enabling pseudonymous communication; the IP address of a user is revealed only to a user's trusted friends.

\paragraphb{Social networks and routing:}
The benefits of using social network links for overlay routing has been 
recognized in a large body of academic work as well as deployed systems.
\begin{ndss}
Some systems maintain traditional peer-to-peer structures but also make use of social network
connections, to improve security of the DHT (Sprout~\cite{sprout}),
to improve %
user 
privacy (OneSwarm~\cite{oneswarm}), increase
download speeds in BitTorrent (Tribler~\cite{tribler}), and assist
in peer discovery (Maze~\cite{maze}). These systems are not resilient
to Sybil attacks, and may require flooding, harming scalability
to large networks. Moreover, %
they allow  direct contacts
over untrusted links, 
exposing 
users' IP addresses. %
\end{ndss}

\begin{techreport}
\paragraphe{Hybrid routing using social network links:} 
Systems in this class 
maintain traditional peer-to-peer structures but also make use of social network 
connections. Sprout~\cite{sprout} proposed augmenting the finger tables in 
traditional DHTs, such as Chord, with social network links. The authors showed 
that the added connections could improve the security of the routing mechanism.
However, Sprout does not defend against Sybil attacks, and is not concerned with user privacy. 
OneSwarm~\cite{oneswarm} is a deployed peer-to-peer communication system for improving 
user privacy where routing is performed by combining trusted and untrusted peer relationships.
Tribler~\cite{tribler} increases download speed in BitTorrent by discovering and
downloading file chunks stored at peers. 
Similarly, Maze~\cite{maze} leverages a social network to discover peers and cooperatively download files. 
These three systems leverage flooding to provide
any-to-any reachability, and thus cannot scale to large networks.
The hybrid systems are not resilient to Sybil attacks.  Moreover, %
they allow direct contacts 
over untrusted links, 
exposing
users' IP addresses. %
\end{techreport}

\begin{ndss}
Other systems constrain routing to social network links, to improve privacy. 
WASTE~\cite{waste},Turtle~\cite{turtle},Freenet~\cite{freenet} and MCON~\cite{membership-concealing} are examples of such \emph{darknets}.
Membership concealing overlay networks (MCONs) (formalized by Vasserman et al.~\cite{membership-concealing}), 
hide the real-world identities of the participants through the use of overlay and DHT-based routing.
However, their design makes use of a trusted centralized server and also requires flooding when
a new user joins the network. 
In addition to these limitations, none of the above systems are resilient to Sybil attacks.
\end{ndss}
\begin{techreport}
\paragraphe{Routing only using social network links:} 
All communication in this class of systems is over social network links. This enables 
participants in the network to be hidden from each other, providing a high degree of privacy. Such a network is commonly known as a \emph{darknet}. 
WASTE~\cite{waste} is a deployed decentralized chat, instant messaging, and file sharing protocol, and is widely considered to be the first darknet. 
WASTE does not attempt to scale beyond small networks, and its suggested size is limited to 50 users.
Turtle~\cite{turtle} is a deployed decentralized anonymous peer-to-peer communication protocol. 
Nodes in Turtle do not maintain any state information other than their trusted friend links and use controlled flooding to search 
for data items. Flooding methods \matt{create significant overhead as network size increases.}
Freenet~\cite{freenet} is a deployed decentralized censorship-resistant distributed storage system. 
Version $0.7$ of Freenet nodes can be configured to run in darknet or opennet mode; the latter allows connections from untrusted nodes, and is expected to be used by less privacy-sensitive users.
Freenet's routing algorithm is heuristic and does not guarantee that data will be found at all; it has also been shown to be extremely vulnerable even against a few malicious nodes~\cite{pitch-black}.
Membership concealing overlay networks (MCONs) (formalized by Vasserman et al.~\cite{membership-concealing}), 
hide the real-world identities of the participants through the use of overlay and DHT-based routing.
However, their design makes use of a trusted centralized server and also requires flooding when
a new user joins the network. 
In addition to these limitations, none of the above systems are resilient to Sybil attacks.
\end{techreport}

\section{Limitations}
\label{sec:limitations}

We now discuss some limitations of our design. First, X-Vine 
requires a user's social contacts to be part of the overlay%
\prateek{; the DHT needs to be bootstrapped from a single contiguous trust network.}
\prateek{Next, X-Vine assumes that Sybil identities are distributed randomly in the DHT 
identifier space. We emphasize that this assumption is shared by prior systems~\cite{srdhtr}, 
and that defending multi-hop DHTs against targeted clustering attacks is currently an open problem. 
In future work, we will investigate the possibility of adapting the cuckoo hashing mechanism~\cite{sybil-proof-dht} 
proposed by Lesniewski-Laas (for one-hop DHTs) in the context of securing multi-hop DHTs.
X-Vine also does not defend against attackers who target users by compromising nodes close 
to them in the social network topology.}
Finally, applications using X-Vine experience higher than usual latencies since 
all communications are pseudonymous and traverse multiple social network links. 

\section{Conclusions}
\label{sec:conclusion}

We describe X-Vine, a protection mechanism for DHTs 
that operates entirely by communicating over social network links. 
X-Vine requires $O(\log n)$ state, two orders of magnitude less 
in practical settings as compared with existing techniques,
making it particularly suitable for large-scale and dynamic environments.
X-Vine also enhances privacy by not revealing social relationship information %
and by providing a basis for pseudonymous communication.

\begin{ndss}
\footnotesize
\bibliography{paper}
\bibliographystyle{abbrv}
\end{ndss}

\begin{techreport}
\bibliography{paper}
\bibliographystyle{abbrv}
\end{techreport}

\begin{ndss}
\footnotesize
\appendix
\input{appendix}
\end{ndss}

\begin{techreport}
\appendix
\section{Mathematical Analysis of X-Vine:}
\label{sec:appendix}

As a first step in formally modeling X-Vine security, we develop an 
analytic model for routing in X-Vine. The model enhances our
understanding of the relationship between operational 
parameters of X-Vine, and can serve as a stepping stone 
for a complete formal model to analyze X-Vine's security against Sybil identities. 

Let there be $N$ nodes in the system with node identifiers ranging from $0..N-1$. Let $L(0,w)$ be 
the expected lookup path length between the node with identifier $0$ and $w$. Let us suppose that 
node maintain trails with a single successor. Without
loss of generality, the average lookup path length can be computed as follows:
 
{\small
 \begin{equation}
 E(L)=\frac{\sum_{w=0}^{w=N-1} L(0,w)}{N}
 \end{equation}
 }
 
 In the simplest case, $L(0,0)=0$. Let us first compute $L(0,1)$. Note that node $0$ and node $1$ 
 have a {\it trail} between them because they are overlay neighbors. Let $d$ be the average node
 degree in the underlying topology. We assume that the length of the trail between overlay neighbors 
 is close to their shortest path in the social network topology (approximately $\log_d(N)$).
 The lookup will proceed from node $0$ along the trail to node $1$. Thus we have that:
 
{\small
\begin{subequations}
\begin{align}
L(0,1) & =  \text{Expected trail length}  \\
L(0,1) & =  \log_d(N) 
\end{align}
\end{subequations}
}
 
 Notice that there cannot be any shortcutting in the intermediate nodes on the trail path from
 node $0$ to node $1$ because we have assumed the trail to be the shortest path in the social network topology.  
 Let us now compute $L(0,2)$. There are two possible scenarios. In the first case, there may be a trail 
 with an end point at node $2$ {\it going through} node $0$. In this case, the packet is routed along the trail 
 to node $2$. Again, there cannot be any shortcutting along this trail because it is the shortest path. 
 The mean path length in this case is $\frac{\log_d{N}}{2}$. In the second case, the packet will be routed towards 
 overlay node $1$ (successor of node $0$). Thus we have that: 
 
\begin{subequations}
{\small
 \begin{multline}
 L(0,2)   =   P(\text{trail}) \cdot \frac{\log_d N}{2} \\ 
 + (1-P(\text{trail}))\cdot \left(1+l((\log_d N)-1,1,2)\right) 
 \end{multline}
}
 
where $l(x,y,z)$ is defined as the expected path length when
 $z$ is the destination identifier,  $y$ is the next overlay hop in the lookup, and 
 $x$ is the physical distance along a trail between the current node and $y$ (Figure~\ref{fig:vrr}).
This means that $l((\log_d N)-1,1,2)$ is the
expected path length when the destination identifier is 2, the next overlay hop is 1, and the next overlay hop 
is $\log_d N$ hops away from the current node. 
 
\begin{figure}[!t]
\centering
\includegraphics[width=2in]{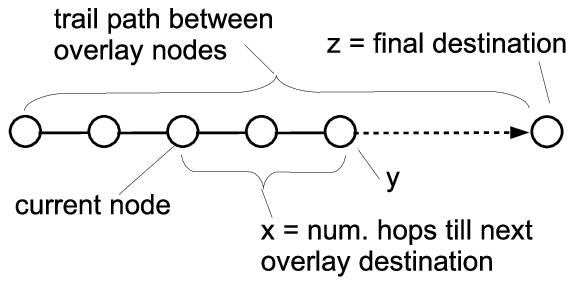}
\caption{X-Vine lookup}
\label{fig:vrr}
\vspace{-0.2in}
\end{figure}
 
Note that each node maintains a trail to its successor, and the mean length of the trails is $\log_d(N)$. 
This means that the average routing state maintained by a node in the system is $\log_d(N)$. Since each routing   
table entry specifies two end points for a trail, the probability that a node has a trail with a particular end point going 
through it is $\frac{2\log_d N}{N}$. Thus:
 
{\small
\begin{multline}
L(0,2)  = \frac{2\log_d N}{N} \cdot \frac{\log_d N }{2} \\ + \left(1-\frac{2\log_d N}{N}\right) \cdot \left(1+l((log_d N)-1,1,2)\right)
\end{multline}
}
\end{subequations}
 
 We now need to compute $l(x,1,2)$. Similar to our computation of $L(0,2)$, again, there are three possible        scenarios. 
 In the first case, the current node (say $A$) is a friend of node $2$. Then the path length is $1$. In the second case, 
 there is a trail with an end point at node $2$ going through node $A$. In this case, the mean path length 
 is $\frac{\log_d(N)}{2}$. In the third case, the packet continues along the current trail to node $1$. 

 {\small
 \begin{multline}
 l(x,1,2) = \frac{2\log_d N}{N} \cdot \left( \frac{\log_d N}{2}\right) \\ +\left(1-\frac{2\log_d N}{N}\right) \cdot (1+l(x-1,1,2))
 \end{multline}
 }

 Here, the boundary conditions for terminating the recursion are as follows: 
 
 {\small
 \begin{subequations}\label{termination}
 \begin{align}
 l(x,1,1) & = x  \text{ if } 0 \leq x \leq \log_d N \\
 l(x,z,z) & = x  \text{ if } 0 \leq x \leq \log_d N, 1 \leq z \leq N-1\\
 l(0,y,z) & = L(y,z) = L(0,(z-y)) \text{ if } 1 \leq y \leq z \leq N-1 
 \end{align}
 \end{subequations}
 }
 
 Let us now compute $L(0,w)$. Consider the following two scenarios. In the first case, let the closest 
 preceding node in node $0$'s routing table be node $i$ (shortcut to $i\neq 1$). Now node $i$ may either 
 be a friend of node $0$, in which case, the path length is $1+L(i,w)$, or node $i$ may be the end point of a 
 trail going through node $0$, in which case, the path length is $1+l\left(\frac{\log_d N}{2}-1,i,w\right)$. 
 In the second case, there is no shortcutting, and the lookup proceeds towards the next overlay hop node $1$. 
 Thus, we have that:

 {\small
 \begin{multline}\label{main-1}
 L(0,w)=  \sum_{i=2}^{w} P(\text{shortcut to i}) \cdot   P(\text{shortcut via friend}) \\ \cdot (1+L(i,w)) 
    + \sum_{i=2}^{w} P(\text{shortcut to i}) \cdot \\ P(\text{shortcut via trail}) \cdot  \left(1+l\left(\frac{\log_dN }{2}-1,i,w\right)\right) \\
    + P(\text{no shortcut}) \cdot  (1+l((\log_d N)-1,1,w)) 
 \end{multline}
 }
 
 Let us now compute the probability of shortcutting to a node $i$. The probability of shortcutting to 
 node $w$ is simply $\frac{d+2\log_d N}{N}$. The probability of shortcutting to node $w-1$ can be computed 
 as $P(\text{no shortcut to }w) \cdot P(\text{shortcut to }w-1|\text{ no shortcut to }w)$. This is
 equal to $\left(1-\frac{d+2\log_d N}{N}\right)\cdot\frac{d+2\log_d N}{N-1}$. Similarly, we can compute the probability
 of shortcutting to node $i$ as:

 {\small
 \begin{subequations}\label{short}
 \begin{align}
 P(\text{shortcut to i}) & =  P(\text{no shortcut to w..i+1}) \cdot  \nonumber \\ \frac{d+2\log_d N}{N-(w-i)} \\ 
 P(\text{no shortcut to w..j}) &= P(\text{no shortcut from w..j+1}) \cdot \nonumber \\  \left(1-\frac{d+2\log_d N}{N-(w-j)}\right) 
 \end{align}
 \end{subequations}
 }
 
 Now, given that the lookup shortcuts towards overlay hop node $i$, it may do so because of a friendship
 entry in the routing table, or a trail in the routing table. The probability that the shortcut happened
 via a friend entry, $P(\text{shortcut via friend})=\frac{d}{d+2\log_d(N)}$. The probability that the          shortcut 
 happened because of a X-Vine entry is \\ 
$P(\text{shortcut via trail})=\frac{2\log_d(N)}{d+2\log_d(N)}$. 
Thus,   we can rewrite
 equation~\eqref{main-1} as
 
 {\small
 \begin{multline}
 L(0,w)= \sum_{i=2}^{w} P(\text{shortcut to i}) \cdot \frac{d}{d+2\log_d N} \cdot (1+L(i,w)) \\ 
  + \sum_{i=2}^{w} P(\text{shortcut to i}) \cdot \frac{2\log_d N}{d+2\log_d N} \cdot \left(1+                    l\left(\frac{\log_d N}{2}-1,i,w\right)\right)\\
   +P(\text{no shortcutting}) \cdot (1+l((\log_d N)-1,1,w))
 \end{multline}
 }
 
 Similar to the above analysis, we can compute $l(x,i,w)$ as follows:
 
 {\small
 \begin{multline}\label{main-2}
 l(x,j,w) = \sum_{i=2}^{j+1} P(\text{shortcut to i}) \cdot \frac{d}{d+2\log_d N} \cdot (1+L(i,w)) \\
     + \sum_{i=2}^{j+1} P(\text{shortcut to i}) \cdot \frac{2\log_d N}{d+2\log_d(N)} \cdot \left(1+                l\left(\frac{\log_d N}{2}-1,i,w\right)\right) \\
     + P(\text{no shortcutting} ) \cdot (1+l(x-1,j,w))
 \end{multline}
 }
 
 The boundary conditions for the termination of recursion are the same as in 
 equation~\eqref{termination}.

\paragraphe{Validation of analytic model:} Figure~\ref{fig:pathlength-d10} plots the mean 
lookup path length as a function of number of nodes for a synthetic scale-free topology with 
 average degree $d=10$ using a redundancy parameter of $r=1$. We can see that the results of  simulation are 
a very close match with our analytic model, %
increasing confidence in our results. %
We note that our analytic model has implications for modeling network layer DHTs like VRR. 

\begin{figure}[!t]
\centering
\includegraphics[height=2.3in]{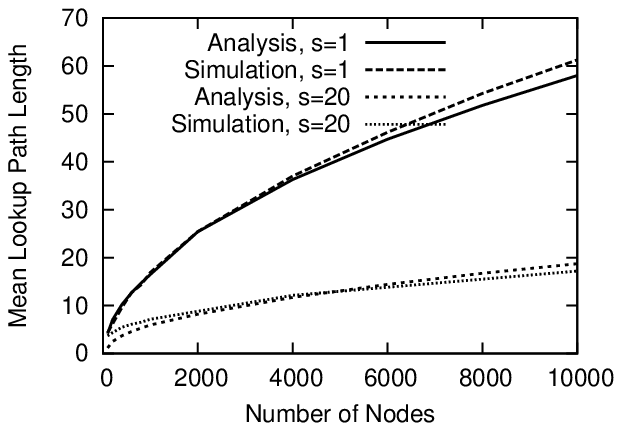}
\caption{Validation of Analytic Model using $d=10$}
\label{fig:pathlength-d10}
\end{figure}

\newpage

\section{Pseudocode}
\label{sec:pseudocode}
\begin{algorithm}
\caption{{\em Fwd\_lookup(identifier myid, message M)}: Determines next hop for a lookup message.}
\label{alg:fwdlookup}
\begin{algorithmic}
\STATE bestroute=0
\STATE {\bf foreach} element E in RoutingTable
\STATE \hspace{0.1in} {\bf if} distance(E.endpoint,M.dest)$<$ 
\STATE \hspace{1in}distance(bestroute,M.dest)
\STATE \hspace{0.1in} \hspace{0.1in} bestroute=E
\STATE {\bf endfor}
\STATE {\bf return} bestroute
\end{algorithmic}
\end{algorithm}

\begin{algorithm}
\caption{{\em Fwd\_trailsetup(identifier myid, message M):} Determines next hop for trail path setup message.}
\label{alg:fwdtrailsetup}
\begin{algorithmic}
\STATE bestroutes=$\emptyset$
\STATE /* select all routes that make progress */
\STATE {\bf foreach} element E in RoutingTable
\STATE \hspace{0.1in} {\bf if} distance(E.endpoint,M.dest)$<$distance(myid,M.dest)
\STATE \hspace{0.1in} \hspace{0.1in} bestroutes.insert(E)
\STATE {\bf endfor}
\STATE /* of these, discard (a) backtracked routes, (b) routes that have reached bounds, (c) routes that don't make namespace progress compared to M.nextovlhop*/
\STATE {\bf foreach} element E in bestroutes
\STATE \hspace{0.1in} {\bf if} failed\_set.contains(E.endpoint,E.nexthop) {\bf or}
\STATE \hspace{1in} (E.nexthop.numtrails $>$ $b_n$) {\bf or} 
\STATE \hspace{1in} (numtrailsto(E.nexthop) $>$ $b_l$) {\bf or}
\STATE \hspace{1in} (distance(E.endpoint,M.dest) $<$ 
\STATE \hspace{1.2in}distance(M.nextovlhop,M.dest)
\STATE \hspace{0.1in} \hspace{0.1in} bestroutes.remove(E)
\STATE {\bf endfor}
\STATE /* if no remaining options, backtrack */
\STATE {\bf if} bestroutes == $\emptyset$
\STATE \hspace{0.1in} send\_reject\_to(M.prevhop)
\STATE \hspace{0.1in} {\bf return}
\STATE /* of remaining routes, select route with maximum namespace progress */
\STATE routetouse=0
\STATE {\bf foreach} element E in bestroutes
\STATE \hspace{0.1in} {\bf if} distance(E.endpoint,M.dest)$<$ 
\STATE \hspace{1in}distance(routetouse,M.dest)
\STATE \hspace{0.1in} \hspace{0.1in} routetouse=E
\STATE {\bf endfor}
\STATE {\bf return} routetouse
\end{algorithmic}
\end{algorithm}

\end{techreport}

\end{document}